%% file: 00_Main.tex
\documentclass[conference]{IEEEtran}
\IEEEoverridecommandlockouts
\usepackage{cite}
\usepackage{amsmath,amssymb,amsfonts}
\usepackage{algorithmic}
\usepackage{graphicx}
\usepackage{textcomp}
\usepackage{amsthm}
\usepackage{xcolor}
\usepackage{bbding}
\usepackage{makecell}
\usepackage{algorithm}  
\usepackage{color}
\usepackage{multirow}
\usepackage{bm}
\usepackage{array}
\usepackage{booktabs}
\usepackage{tabularx,booktabs}
\usepackage{multicol}
\usepackage{cleveref}
\usepackage{balance}
\usepackage{lastpage}
\usepackage{tabulary}
\usepackage{subfigure}
\usepackage{etoolbox}
\usepackage{bbm}
\usepackage{fancyhdr}
\usepackage{enumitem}
\usepackage{mathrsfs}
\usepackage{multicol}
\usepackage{amsfonts,amssymb}
\usepackage{ulem}
\usepackage{cancel}
\usepackage{setspace}
\usepackage{stackengine}
\usepackage{threeparttable}

\usepackage{amsthm}
\theoremstyle{definition}

\usepackage{nomencl}
\makenomenclature

\def\BibTeX{{\rm B\kern-.05em{\sc i\kern-.025em b}\kern-.08em
    T\kern-.1667em\lower.7ex\hbox{E}\kern-.125emX}}

\newcommand{\blue}[1]{\textcolor{black}{#1}}

\begin{document}

\bstctlcite{IEEEexample:BSTcontrol}

\title{Large Language Models for Wireless Networks: An Overview from the Prompt Engineering Perspective\\
\thanks{Hao Zhou, Chengming Hu, Dun Yuan, Ye Yuan, Xi Chen, and Xue Liu are with the School of Computer Science, McGill University, Montreal, QC H3A 0E9, Canada. (emails:\{hao.zhou4, chengming.hu, dun.yuan, ye.yuan3\}@mail.mcgill.ca, xi.chen11@mcgill.ca, xueliu@cs.mcgill.ca); 
Di Wu is with the School of Electrical and Computer Engineering, McGill University, Montreal, QC H3A 0E9, Canada. (email: di.wu5@mcgill.ca);
Hina Tabassum is with the Department of Electrical Engineering and Computer Science at York University, Toronto, ON M3J 1P3, Canada. (e-mail:hinat@yorku.ca).}}
\author{\IEEEauthorblockN{Hao Zhou, Chengming Hu, Dun Yuan, Ye Yuan, Di Wu, Xi Chen,\\ 
Hina Tabassum, \IEEEmembership{Senior member, IEEE},
Xue Liu, \IEEEmembership{Fellow, IEEE}}}

\maketitle

\thispagestyle{fancy}            
\fancyhead[C] {This paper has been accepted by IEEE Wireless Communications Magazine. } 

\begin{abstract}
Recently, large language models (LLMs) have been successfully applied to many fields, showing outstanding comprehension and reasoning capabilities. 
Despite their great potential, LLMs usually require dedicated pre-training and fine-tuning for domain-specific applications such as wireless networks. 
These adaptations can be extremely demanding for computational resources and datasets, while most network devices have limited computation power, and there are a limited number of high-quality networking datasets.
To this end, this work explores LLM-enabled wireless networks from the prompt engineering perspective, i.e., designing prompts to guide LLMs to generate desired output without updating LLM parameters.
Compared with other LLM-driven methods, prompt engineering can better align with the demands of wireless network devices, e.g., higher deployment flexibility, rapid response time, and lower requirements on computation power.    
In particular, this work first introduces LLM fundamentals and compares different prompting techniques such as in-context learning, chain-of-thought, and self-refinement. Then we propose two novel prompting schemes for network applications: iterative prompting for network optimization, and self-refined prompting for network prediction.  
The case studies show that the proposed schemes can achieve comparable performance as conventional machine learning techniques, and our proposed prompting-based methods avoid the complexity of dedicated model training and fine-tuning, which is one of the key bottlenecks of existing machine learning techniques.   
\end{abstract}


\section{Introduction}

\begin{table*}[!t]
\caption{Summary of various LLM usage approaches for wireless networks.}
\centering
\small
\setstretch{1.1}
\begin{threeparttable}
\resizebox{1\textwidth}{!}{%
\begin{tabular}{|m{1.5cm}<{\centering}|m{3.7cm}<{\centering}|m{3.5cm}<{\centering}|m{3.5cm}<{\centering}|m{4cm}<{\centering}|m{4cm}<{\centering}|}
\hline
Approaches & Key principles & Advantages & Potential difficulties & Computational resources \& Time costs & Possible wireless network applications\\
\hline
Pre-training LLMs 
& 1) Training LLMs from scratch on \blue{hundreds of billions of tokens for the next token prediction.} \qquad
2) Pre-training is a foundational step of LLM development and usage. 
& Pre-training enables LLMs \blue{with comprehension, reasoning, and instruction following capabilities}, which can address various downstream tasks. 
& Pre-training is extremely resource-intensive, requiring long training times, and \blue{substantial energy consumption and GPUs or TPUs}. 
& High Thermal Design Power GPUs such NVIDIA H$100$-$80$G or A$100$-$80$G are required, \blue{e.g., training Llama$3.1$-$405$B took approximately $30.84$M GPU hours.}
& A network-specific LLM indicates great benefits, but training network LLMs from scratch may be inappropriate due to the computational resource requirements. \\
\hline
Fine-tuning & 
Fine-tuning refers to adapting a pre-trained LLM to specific tasks by partially or completely updating the model weights or introducing additional adapter modules. 
& Fine-tuning significantly reduces time and computational cost compared to pre-training, producing LLM models tailored to specific tasks. 
& 1) Fine-tuning has to consider the risk of overfitting and losing generalization. \qquad \qquad
2) Labelled data may be needed \blue{with additional human labour}.
& The cost may vary between model sizes and fine-tuning methods. If full fine-tuning a 7B parameter model with 8 A$100$-$80$G GPUs, it could be $100$-$200$ GPU hours on a $100$M token dataset. 
& Fine-tuning is a realistic approach to apply LLMs to wireless networks, adapting a general-domain LLM to specific tasks, \blue{e.g., network troubleshooting and configuration.}  
\\
\hline
Retrieval augmented generation (RAG)
& RAG combines LLMs with an external knowledge base, enabling the model to retrieve relevant information during inference \blue{and provide more accurate responses}.
& RAG can improve LLM's response accuracy by leveraging external knowledge sources, \blue{generating more up-to-date results}.
& 1) RAG involves extra integration complexity, as it requires efficient retrieval algorithms. \qquad 
2) RAG highly depends on a well-structured knowledge base, requiring dedicated creation.
& 1) The cost depends on the retrieval method and the size of the knowledge base. Inference may take seconds to minutes per query. \qquad \qquad
2) Requires a balance between retrieval speed and accuracy.
& 1) RAG is a more practical approach for implementing LLMs, especially when existing network-specific datasets are available. \qquad \quad 
2) It ensures the LLM has access to the most relevant and up-to-date network data. 
\\
\hline
Prompt engineering 
& Prompt engineering refers to crafting specific inputs to guide a pre-trained LLM to generate desired outputs, without any model retraining or tuning.
& 1) Prompt engineering has low computational costs and quick implementation. 2) It can be easily adapted to different tasks without additional training.
& 1) \blue{It requires expertise in crafting effective prompts.}
2) It is also limited by the inherent capabilities and context window size of the pre-trained LLMs.
& \blue{1) Without the need for backward passes and gradient updates, it requires minimal computational resources.} \qquad
2) Time costs are significantly lower compared to other methods.
& 
1) Prompt engineering enables rapid decision-making and responses in dynamic wireless network environments. \qquad \quad
2) It aligns well with many network tasks.\\
\hline
\end{tabular}}
\label{tab-usage}
\begin{tablenotes}    
        \footnotesize       
        \item[1] \blue{Note that the objective of this table is to introduce the features of each technique, and they serve different purposes in LLM development and application.\\   
        These approaches can be combined to achieve better performance, e.g., fine-tuning network LLMs and then prompting.}  
\end{tablenotes} 
\end{threeparttable}  
\end{table*}

As a sub-field of generative AI, large language models (LLMs) have received considerable interest from industry and academia \cite{zhou2024large2}.
%
The advancement of generative AI and LLMs also provides promising opportunities for 6G networks, including strong reasoning and planning capabilities, multi-modal understanding for 6G sensing, semantic communication~\cite{liang2023generative}, integrated satellite-aerial-terrestrial networks~\cite{javaid2024leveraging}, etc. 
Despite the great potential, integrating LLMs into wireless networks still faces several challenges.
Firstly, wireless networks are complex large-scale systems with various knowledge domains, i.e., signal processing and transmission, network architecture and design, protocol, standards, and so on. Applying general-domain LLMs directly to domain-specific network tasks may lead to poor performance.   
Secondly, the development of LLMs relies on high-quality datasets for fine-tuning adaptation, while there is a limited amount of high-quality networking datasets such as SPEC5G and Tspec-LLM \cite{zhou2024large2}.
Moreover, LLMs are extremely demanding in terms of computational resources. LLM pre-training and fine-tuning are usually implemented on high-performance GPUs such as NVIDIA A100 and H100, but wireless network devices usually have limited computational and storage capacities.
\blue{LLMs involve a broad range of techniques, such as pre-training, fine-tuning LLMs for domain-specific tasks, retrieval augmented generation (RAG), prompt engineering, etc. 
Therefore, it is critical to identify an efficient method to better adapt LLMs to wireless networks.}

Given the above opportunities and challenges, this work \blue{introduces} prompt engineering, which is regarded \blue{as a resource-efficient and flexible approach to using LLMs with fast implementation speed\blue{\cite{sahoo2024systematic}}}. 
\blue{These advantages will help to overcome the above LLM application challenges, e.g., deployment difficulties and requirements for computational resources.}
In particular, prompting refers to designing input prompts to guide pre-trained LLMs to generate desired outputs. It takes advantage of the inherent inference capabilities of pre-trained LLMs, and avoids the \blue{need for backward passes and gradient updates}.  
\blue{Therefore, prompt engineering has several key features:} 1) \textbf{Resource-efficient}: Prompting only needs forward passing of the model, and no need to store all intermediate activations for backpropagation. Such a resource-efficient approach \blue{may mitigate the computational burden of network servers and devices}.
2) \textbf{Higher flexibility}: Prompting-based methods can quickly adapt to various tasks by crafting the corresponding textual demonstrations and queries without extra coding steps\blue{\cite{du2024generative}}. It indicates an efficient method to customize LLMs to address a wide range of network tasks. 
3) \textbf{Fast implementation}: Prompt engineering relies on the inference capabilities of LLMs, and it avoids the time cost of updating LLM parameters. Therefore, the low response time can better handle low-latency network services.    
\blue{Finally, note that although prompting LLMs can significantly enhance the performance, it is not a standalone solution for LLM usage\cite{zhao2023survey}. For instance, combining prompt engineering with fine-tuning may create a robust framework with both prompting flexibility and model specialization. However, this work focuses on prompting techniques to investigate their full potential.}

LLMs have been discussed in several existing studies, but they mainly focus on system-level discussions and module designs, e.g., edge intelligence \cite{lin2023pushing} and grounding and alignment\cite{xu2024large}. Prompting engineering is also used for AI-generated everything services in \cite{liu2023optimizing}, and our previous work also investigates LLM-enabled power control and traffic prediction in \cite{zhou2024large, hu2024self}.
However, this work is different from existing studies by systematically exploring prompt engineering and wireless network applications, providing detailed prompt designs and specific case studies. 
The main contributions are:

1) Firstly, we present in-depth analyses of LLM fundamentals and the feasibility of network applications such as pre-training, fine-tuning, RAG, and prompt engineering. 
\blue{Each technique is introduced} in terms of computational resources and time costs, potential difficulties, and wireless network applications.  
Then, we provide an overview of a variety of prompting techniques and discuss their applications to wireless networks, such as in-context learning, chain-of-thought, prompt-based planning, and self-refinement.

2) We propose two novel prompting techniques, namely iterative prompting and self-refined prompting, aiming to address network optimization and prediction problems, respectively.
Specifically, iterative prompting learns from previous experiences to improve LLM's performance on target tasks iteratively, which is suitable for handling network optimization problems.
By contrast, self-refined prompting can correct their outputs
through iterative feedback and refinement prompts, aiming to address network prediction tasks. 
These schemes rely on LLM's inference capabilities, and such resource-efficient techniques align well with the limited computational capacities of many network devices. 
In addition, our proposed algorithms can be easily generalized to various wireless network applications, and the case studies show that they achieve satisfactory performance in network power control and traffic prediction problems. 

\begin{figure*}[t]
\centering
\includegraphics[width=0.95\linewidth]{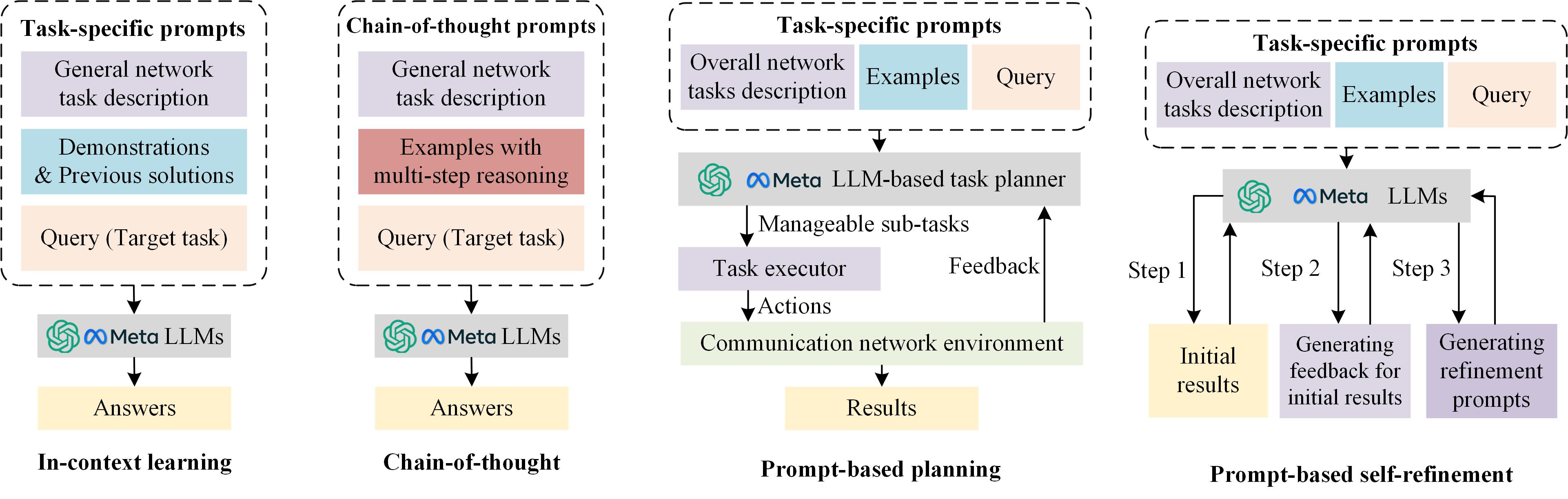}
\caption{Illustration of different prompting techniques.}
\label{fig-prompt}
\end{figure*}

\input{01_overview}

\input{02_Prompting}

\input{03_optimization}

\input{04_prediction}

\input{05_case}

\section{\blue{Real-world Challenges and Limitations}}

\blue{This section will further discuss the limitations and real-world challenges of prompt engineering applications.} 

\blue{Firstly, prompt engineering is inherently limited by the context window size of LLMs, which restricts the capabilities to process tasks requiring extensive contextual information beyond the model's input and memory constraints.
Recent advancements such as Llama3 with a context window size of 128k tokens, significantly mitigate this limitation.}

\blue{Secondly, the effectiveness of prompt engineering relies heavily on the selection of demonstrations shown to the model or the order of demonstrations.
Poorly chosen or suboptimal examples can result in biased outputs.
Our work considers iterative prompting to select better demonstrations, but professional experiences are still needed in this process.}

\blue{Moreover, crafting effective prompts for complicated tasks in network environments usually requires trial-and-error experimentation. This iterative process is inefficient, especially when results are inconsistent or difficult to reproduce. 
Without clear guidelines, engineers must rely on intuition or extensive testing to achieve desired outcomes. This can slow down the development for novel and complex use cases as in 6G.}

\blue{In addition, in real-world applications, prompt engineering may introduce significant security concerns, as it is potentially vulnerable to attacks such as prompt injection and leakage. 
Such attacks are particularly concerning in sensitive domains like wireless communications, where malicious actors could manipulate the model or extract encoded sensitive data\cite{du2023spear}.} 

\blue{Finally, note that prompt engineering is not a standalone solution but one part of pipelines of LLM and generative AI techniques \cite{du2024generative}. 
Prompt engineering is constrained by the inherent limitations of the used LLMs. If the LLM lacks sufficient knowledge or context, prompting is unlikely to produce accurate results. 
Therefore, when prompt engineering struggles with highly specific tasks, it may require external support such as fine-tuning. This dependency highlights that prompt engineering can be actively combined with other techniques such as pre-training and fine-tuning to make the most of LLM potential.}

\section{Conclusion}
The progress of generative AI and LLMs brings promising opportunities for next-generation wireless networks. This work provides a comprehensive overview of prompt engineering for network applications. 
It presents in-depth analyses of different prompting techniques such as in-context learning, chain-of-thought, and self-refinement.
In addition, this work proposes two novel prompting schemes for network optimization and prediction problems, and the case studies show that the proposed methods can achieve satisfactory performance.
\blue{In the future, we will explore more complicated and advanced network tasks such as reconfigurable intelligence surfaces and integrated sensing and communication, uncovering the full potential of LLMs for future wireless networks.}

\normalem
\bibliographystyle{IEEEtran}
\bibliography{Reference}

\end{document}

%% file: 01_overview.tex
\section{LLM Fundamentals towards Wireless Network Applications}
\label{sec-fundamental}

This section introduces various approaches for \blue{integrating} LLMs to wireless networks, including pre-training, fine-tuning, RAG, and prompt engineering. 
Specifically, Table \ref{tab-usage} \blue{presents} these techniques in terms of their working principle, advantages, potential difficulties, computational resources and time costs, and possible wireless network applications.

Firstly, pre-training is a foundational step for LLM development, which includes data collection, filtering, and model training. Pre-training enables LLMs to develop fundamental capabilities such as comprehension, reasoning, and instruction following.
\blue{Then, these capabilities are the foundation for developing other LLM-inspired techniques such as prompt engineering and planning, e.g., prompt engineering relies on the inherent capabilities of LLMs.}
However, it requires considerable computational resources such as NVIDIA H$100$-$80$G or A$100$-$80$G for implementation, and the training process may take weeks and months.

\blue{Given the training costs, pre-training a wireless-specific LLM from scratch can be extremely time and energy-consuming. By contrast, fine-tuning is a more affordable approach for adapting LLMs to network domains.} 
\blue{Given pre-trained general-domain LLMs}, fine-tuning indicates adapting LLMs to specific tasks by partially updating the model weights to fit smaller and domain-specific datasets such as \blue{3GPP, 5G, and O-RAN standards}.
The fine-tuning costs are related to the LLM model size and fine-tuning methods. For instance, fully fine-tuning a 7B LLM model using 8 A100 GPUs may take 100-200 GPU hours for a 100M token dataset.  
\blue{For large-scale datasets in wireless fields, e.g., SPEC5G dataset with 134M words and 3GPP dataset Tspec-LLM with 534M words \cite{zhou2024large}, the costs on computational resources may be larger.}
Similarly, RAG is another promising direction for practical LLM applications for wireless networks. It integrates LLMs with an external knowledge base, and then the LLM can retrieve related information from the external knowledge source during the inference.
Since LLMs have access to the most relevant and up-to-date information, \blue{e.g., novel network standards, architecture and signal transmission techniques}, RAG can significantly improve the generation quality. 

Instead of updating model parameters, \blue{prompt engineering} utilizes LLM's inference capabilities, designing specific input prompts to guide the generation of pre-trained LLMs.
\blue{Importantly, prompt engineering avoids the need for backward passes and gradient updates of LLMs, and therefore the time costs are significantly lower than fine-tuning.} 
Due to the low costs, it can quickly adapt to dynamic network environments, handling various network tasks with much lower response time. 
%
\blue{In addition, prompting engineering also} aligns well with several crucial wireless network features. 
Firstly, many network devices are computational and energy resource-constrained, \blue{and prompt engineering has low requirements for computational resources, which may mitigate the energy consumption and computational burden on network devices.}
%
Secondly, prompt engineering allows human language-based instructions, e.g., network-specific task descriptions, questions, and solutions. By integrating natural language, network intelligence will become more accessible, especially for operators with minimal professional AI knowledge. 
In addition, prompt engineering has high design flexibility. 
\blue{With proper textual templates, prompts can be efficiently designed using human language and applied to various network tasks, which also have minimum requirements on mathematical equations.} 
\blue{Moreover, prompting can be easily combined with other techniques, e.g., fine-tuning LLMs on network-specific datasets, and then prompting the model to further improve the generated content.}

\blue{On the other hand, prompt engineering also requires considerable expertise in crafting the prompts, which will directly affect the quality of generated content. Although prompt engineering can significantly improve LLM's performance, it relies on the model's inherent capabilities, e.g., achieved by pre-training or fine-tuning.}  
%
Therefore, given the \blue{opportunities and challenges}, it is crucial to explore the integration of prompting techniques and wireless networks, showing a promising way to utilize generative AI.

\begin{table*}[!t]
\caption{Summary of different prompting techniques for wireless networks. \textbf{\blue{\quad(Updated)}} }
\centering
\small
\setstretch{1.1}
\resizebox{1\textwidth}{!}{%
\begin{tabular}{|m{1.5cm}<{\centering}|m{4.5cm}<{\centering}|m{4cm}<{\centering}|m{4.5cm}<{\centering}|m{4.5cm}<{\centering}|}
\hline
Prompting techniques & Key principles & Advantages & Potential issues & Wireless network applications\\
\hline
In-context learning 
& In-context learning empowers LLMs to learn from task-specific instruction and demonstrations. These prompts are designed with task-specific examples that are comprehensible to LLMs.  
& LLMs can be generalized to perform various learning tasks by utilizing existing knowledge from pre-training data and acquiring new task-solving strategies from contextual demonstration prompts. 
& Demonstrations greatly impact in-context learning, which should be carefully designed, \blue{e.g., selection, format, and order of examples.}  
& In-context learning is a promising approach to applying LLMs to wireless networks, using previously accumulated solutions as input prompts to address new network issues. 
\\
\hline
Chain-of-thought  
& 1) Chain-of-thought is an advanced strategy to improve LLM's capabilities in complex reasoning. 2) It can guide LLMs to better understand the logical steps from questions to answers. 
& LLMs can solve complex problems more transparently and logically through chain-of-thought demonstration prompts, \blue{e.g., “\textit{Let's think step by step}”}.
& Chain-of-thought prompting faces issues like incorrect reasoning and instability. Their inherent instability may provide incorrect and misleading intermediate reasoning steps. 
& \blue{Chain-of-thought} can make LLMs better adapt to complicated wireless environments that require \blue{step-by-step reasoning and planning}, e.g., project code generation with multi-step scheduling.\\
\hline 
Prompt-based planning 
& Prompt-based planning aims to address more complicated scenarios such as multi-hop question answering. It breaks complex tasks down into manageable sub-tasks.   
& LLMs can make complex tasks more manageable by using their planning capabilities, lowering the overall difficulties.    
& LLM-based planning requires dedicated analyses to decompose a complicated task into multiple sub-tasks. Zero-shot and automated task decomposition is still the main difficulty. 
& Wireless networks are complicated systems and LLM-enabled automated task decomposition can be extremely useful for long-term network management and project development.   \\
\hline
Self-refinement  
& Self-refinement allows them to correct their outputs through iterative feedback and refining demonstration prompts.  
& Through iterative feedback generation and prediction refinement, LLMs can self-enhance their outputs with strong effectiveness and scalable capabilities. 
& All historical outputs are iteratively appended as the inputs of LLMs, and it may lead to increasingly lengthy demonstration prompts and unavoidable resource consumption. 
& Self-refinement will significantly \blue{save human efforts for network tasks}, since LLMs can constantly improve the previous network policy automatically. 
\\
\hline
\end{tabular}}
\label{tab-prompting}
\end{table*}

%% file: 02_Prompting.tex
\section{Prompt Engineering for Wireless Networks}

\begin{figure*}[t]
\centering
\includegraphics[width=0.9\linewidth]{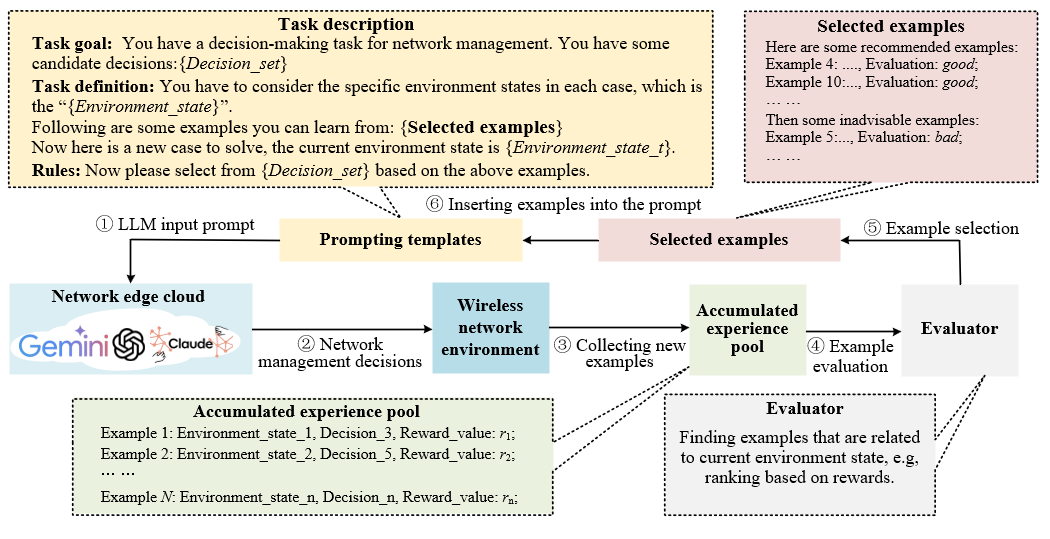}
\caption{\blue{Iterative prompting schemes for wireless network optimization}}
\label{fig-optimization}
\end{figure*}

This section presents several prompting techniques and their network applications, i.e., in-context learning, chain-of-thought, prompt-based planning, and self-refinement methods.

Firstly, as shown in Fig.\ref{fig-prompt}, in-context learning is a fundamental and crucial capability of LLMs. It refers to the process of learning from formatted natural language instructions and demonstrations, and then improving the performance on target tasks.  
Given contextual demonstration prompts, LLMs can address various downstream tasks by using existing knowledge from pre-trained data.
Demonstrations are of great importance for in-context learning, serving as critical references for LLMs to learn from.    
Therefore, they should be carefully selected, formatted, and designed, which may require professional understanding and knowledge. 
In-context learning is a promising approach to applying LLMs to wireless networks, which can take advantage of previous network solutions as demonstrations to address unseen network issues.

Meanwhile, wireless networks are complicated systems, and one task may include several logic and reasoning steps. For instance, many network optimization problems include several elements, e.g., base stations, transmission power, bandwidth, and network users. 
Using simple input-output pairs as demonstrations cannot provide the logical reasons behind these examples, and LLMs may have difficulty in learning. 
To this end, chain-of-thought is proposed as an advanced strategy to enhance LLM's performance in complex reasoning tasks. 
Specifically, chain-of-thought will explain each logic step in the prompt demonstrations: “\textit{Let's think step by step. Given the current base station transmission power $\{BS\_power\}$, using the Shannon capacity equation, the transmission data rate is $\{user\_rate\}$. Then, we compare the achieved rate with the target data rate, and $\{user\_rate\}$ is lower. Therefore, we may need to increase the transmission power.}”  
This example includes two steps, \textit{“using Shannon capacity equation”} and \textit{“compare the achieved rate with the target data rate”}, \blue{and} then the final decision is to \textit{“increase the transmission power”}. With these two-step explanations, LLMs can easily capture the relationship between transmission power and target data rate, and generate better replies using the logical chain.

In-context learning and chain-of-thought are mainly designed for single tasks, which may have difficulty tackling more complicated tasks, e.g., collecting information from multiple sources and producing multi-step reasoning to generate a comprehensive answer. 
Many complex network tasks need to be broken down into more manageable sub-tasks. For example, the network configuration usually involves multiple interrelated network elements, and network project development may consist of a series of coding and testing tasks. 
In this case, prompt-based planning aims to generate a group of sub-tasks, which will be executed one by one. This planning capability is crucial for handling many large-scale network tasks.

In addition, LLMs may generate incorrect answers at the initial attempts, and external feedback is essential to improve its replies. 
As shown in  Fig.\ref{fig-prompt}, self-refinement allows LLMs to self-improve their previous outputs by generating feedback and refinement prompts. 
For instance, the LLM's initial transmission power decision may be incorrect, and we can use the same LLM to evaluate and provide feedback, and then feed the feedback and refinement to the LLM, improving the initial power control decisions. 
In particular, it indicates that LLMs can automatically improve their previous outputs without human intervention.
Self-refinement can significantly save human effort in improving LLM's initial replies, especially considering the complexity of network tasks.

Finally, Table \ref{tab-prompting} compares the above techniques in terms of main features, advantages, potential issues, and wireless network applications.
Note that these methods can be combined to address complex tasks comprehensively. For instance, using self-refinement in prompt-based planning to save effort in the evaluation phase, and integrating chain-of-thought to improve the multi-step logic reasoning in sub-task decomposition.

%% file: 03_optimization.tex
\section{ Iterative Prompting for Wireless Network Optimization}
\label{sec-opti}

Optimization techniques are of great importance for wireless network management, and this section proposes a novel iterative prompting scheme, in which we prompt LLMs iteratively to improve the performance on target network tasks.

\subsection{Natural Language-based Task Description}

Firstly, as shown in Fig. \ref{fig-optimization}, the target task is described by formatted natural language, including task goals, definitions, and extra rules, providing fundamental task information to the LLM. 
For instance,  the task goal describes the problem as a “\textit{decision-making task}”, and it introduces the set of control variables as “\textit{candidate decisions}”.
Then, it highlights the importance of “\textit{environment states}”, indicating that the LLM has to consider specific environment variables for decision-making.
After that, it introduces the current target task by “\textit{here is a new case to solve, and the current environment state is...}”. Moreover, the final rules specify that “\textit{select from \{Decision\_set\} based on above examples}”.

With these formatted task descriptions, the LLM can capture the key elements of the decision-making task, focusing on hidden patterns between environment states and corresponding decision variables.   
Such a design can also lower the understanding difficulties caused by professional network knowledge, since general domain LLMs such as GPTs and Llama are not specifically pre-trained for network applications.

\begin{figure*}[t]
\centering
\includegraphics[width=0.98\linewidth]{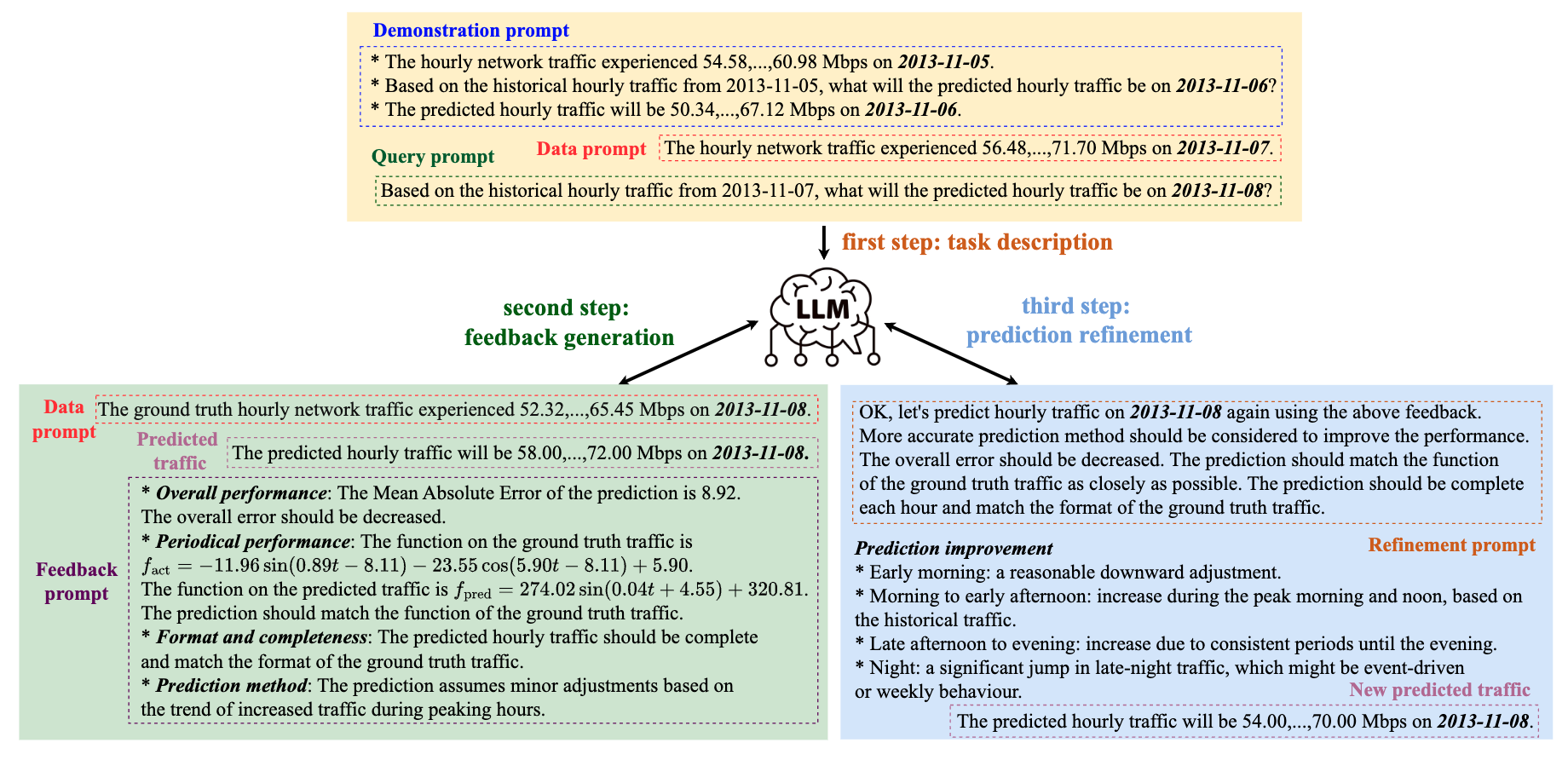}
\caption{\blue{Self-refined prompting for network traffic prediction. 
As an example, by following prediction tasks framed in a question-answer format, the LLM is guided to predict hourly traffic prediction for 2013-11-08, using hourly historical traffic from 2013-11-07.
Due to space constraints,  we show traffic for selected hours only, while hourly traffic is fully utilized during interactions with the LLM. 
The specific feedback related to initial predictions is incorporated into the feedback prompt. The LLM further refines its predictions by following the actionable steps outlined in the refinement prompt.
The process of feedback generation and prediction refinement is iteratively conducted until the prediction performance converges, whereas the inference is completed without engaging in the feedback generation and prediction refinement.}}
\label{fig_self_refine}
\end{figure*}

\subsection{Iterative Prompting Scheme}

As illustrated in Fig. \ref{fig-optimization}, the above task description will become input prompts for the LLM at network edge cloud, and the LLM selects a specific \textit{Decision\_n} from the set of candidate decisions \textit{\{Decision\_set\}}.     
Then, the network management decisions will be sent to the wireless network environment for implementation such as resource allocation, user association and transmission power level decisions.
The network operation results be collected as a new example, including environment states, selected decisions, and rewards that are related to optimization objectives and network metrics. 

After that, the examples from the network output, i.e., previous explorations, are collected in an experience pool, recording all previous examples. The examples in this pool will be further evaluated to choose specific examples well-suited for the target task.
For instance, a straightforward strategy is to select previous examples with similar environment states as current states, e.g., similar user numbers and channel conditions. Then the previous decisions and rewards become useful references for LLM decision-making.   
However, note that many wireless environment states are continuous variables, e.g., channel state information or user-base station distances, and therefore it is unlikely to find examples with exactly the same states. Therefore, the evaluation and selection of examples are crucial in the proposed scheme.

Finally, the selected examples, i.e., recommended and inadvisable demonstrations, are integrated into the task description, providing useful references for LLM decision-making. 
Fig. \ref{fig-optimization} has presented steps 1 to 6 of the proposed iterative prompting scheme, and it can iteratively explore new decisions and accumulate new experiences. 
Therefore, the LLM can constantly learn from previous explorations, and then improve its decisions accordingly. Such an iterative prompting method can rapidly adapt to dynamic wireless environments. %

\blue{Fig. \ref{fig-optimization} applies an exploration-exploitation strategy as in reinforcement learning (RL). 
However, iterative prompting has several advantages over RL approaches: 1) No model parameter updating and fine-tuning is required, which has much lower complexity than existing RL algorithms such as deep reinforcement learning (DRL). 
2) It utilizes human natural language for network management. With minimum requirements on mathematical knowledge and equations, it makes network optimization more accessible than using RL. 
3) LLMs can provide reasonable explanations for their decisions, and help humans understand complicated network systems, which is beyond the capabilities of RL techniques.}  
\blue{Finally, note that our proposed schemes can also be generalized to many other fundamental network optimization tasks, providing a novel LLM-based solution for network control.}

%% file: 04_prediction.tex
\section{Self-refined Prompting for Network Prediction}
\label{sec-pre}

Prediction tasks are critical in wireless networks, using historical network data to predict future trends and behaviours such as network traffic, user demand, channel states, and device status.
To showcase the potential of prompting techniques, this section presents a self-refined prompting method for network prediction problems, i.e., network traffic prediction.

\subsection{Prediction Task Description }

A crucial step in using LLMs for prediction problems is to convert numerical values into natural language sentences. 
Fig.~\ref{fig_self_refine} considers network traffic prediction as an example, and illustrates the demonstration prompt, data prompt, and query prompt.
In particular, the traffic prediction task is framed in a question-answer format, adhering to a template-based description. 
The demonstration prompt is first incorporated to guide the LLM in making traffic predictions by leveraging contextual task descriptions for next-day traffic prediction. 
Then, the LLM predicts future traffic by responding to questions in the query prompt, based on historical traffic data provided in the data prompt, i.e., previous hourly network traffic.

\subsection{Feedback Generation and Prediction Refinement}

Initially, LLMs may generate incorrect predictions, 
and this subsection utilizes self-refined prompting to incorporate feedback demonstrations to provide comprehensive and valuable insights, aiming to improve the initial results~\cite{hu2024self}. 
As shown in Fig.~\ref{fig_self_refine}, the feedback prompt is designed to encompass prediction performance, prediction format and completeness, and prediction method, which are shown as follows:

\begin{itemize}
\item \textbf{Overall performance}: The same LLM evaluates the overall prediction performance, e.g., Mean Absolution Error (MAE) between ground truth and predictions. 
\item \textbf{Periodical performance}: Considering that time-series data can be represented through sine and cosine functions in the real domain~\cite{esling2012time}, the same LLM is tasked with projecting both ground truth and predicted traffic onto these functions, respectively. 
By matching the projection of ground truth, predictions are expected to capture the periodic nature of ground truth accurately, such as fluctuations during peak and off-peak traffic periods.
\item \textbf{Format and completeness}: The predictions should match the format of ground truth as well as be complete for each timestamp, e.g., 24 values for daily traffic prediction. 
\item \textbf{Prediction method}: The prediction method is summarized in the feedback demonstration prompt, which calls for adopting more advanced and accurate methods to enhance performance further. 
To avoid additional computational costs, while the LLM may suggest developing a neural network as a potential improvement method, this recommendation is disregarded. 
\end{itemize}

Given the above instructions, the refinement demonstration prompts aim to provide specific actionable steps associated with feedback demonstration prompts. 
This allows the same LLM to self-refine previous predictions by adhering to the detailed feedback outlined in feedback demonstration prompts. 
Note that the process of feedback generation and prediction refinement is iteratively conducted until the prediction performance converges, whereas the inference is completed without engaging in the feedback generation and prediction refinement. 
Additionally, to prevent the repetition of previous incorrect predictions, the history of predicted traffic, feedback, and refinement demonstration prompts are also included as inputs to the LLM at each iteration.

\blue{In summary, LLMs excel at recognizing and performing new tasks through prompt engineering techniques that utilize contextual information, including task descriptions and demonstration examples. 
Specifically, the proposed self-refined prompting method designs task-specific demonstration, feedback, and refinement prompts, allowing the LLM to iteratively enhance its predictions by adhering to the instructions in these prompts, without requiring additional model fine-tuning or training.  
Compared to the dedicated training of long short-term memory (LSTM) networks, the proposed self-refined prompting method is more computationally efficient with the inference abilities, particularly advantageous for on-device learning on resource-constrained devices in real-world wireless networks.} 
Given that iteratively appending all historical predictions can increase the length of demonstration prompts and potentially lead to higher resource consumption, a future direction for the self-refined prompting method will focus on effectively compressing extensive demonstration prompts, thereby further enhancing resource efficiency.

%% file: 05_case.tex
\begin{figure*}[t]
\centering
\subfigure[\textbf{Optimization Case Study}: improved power consumption of Llamas.]{ \label{fig_power}
\includegraphics[width=0.44\linewidth]{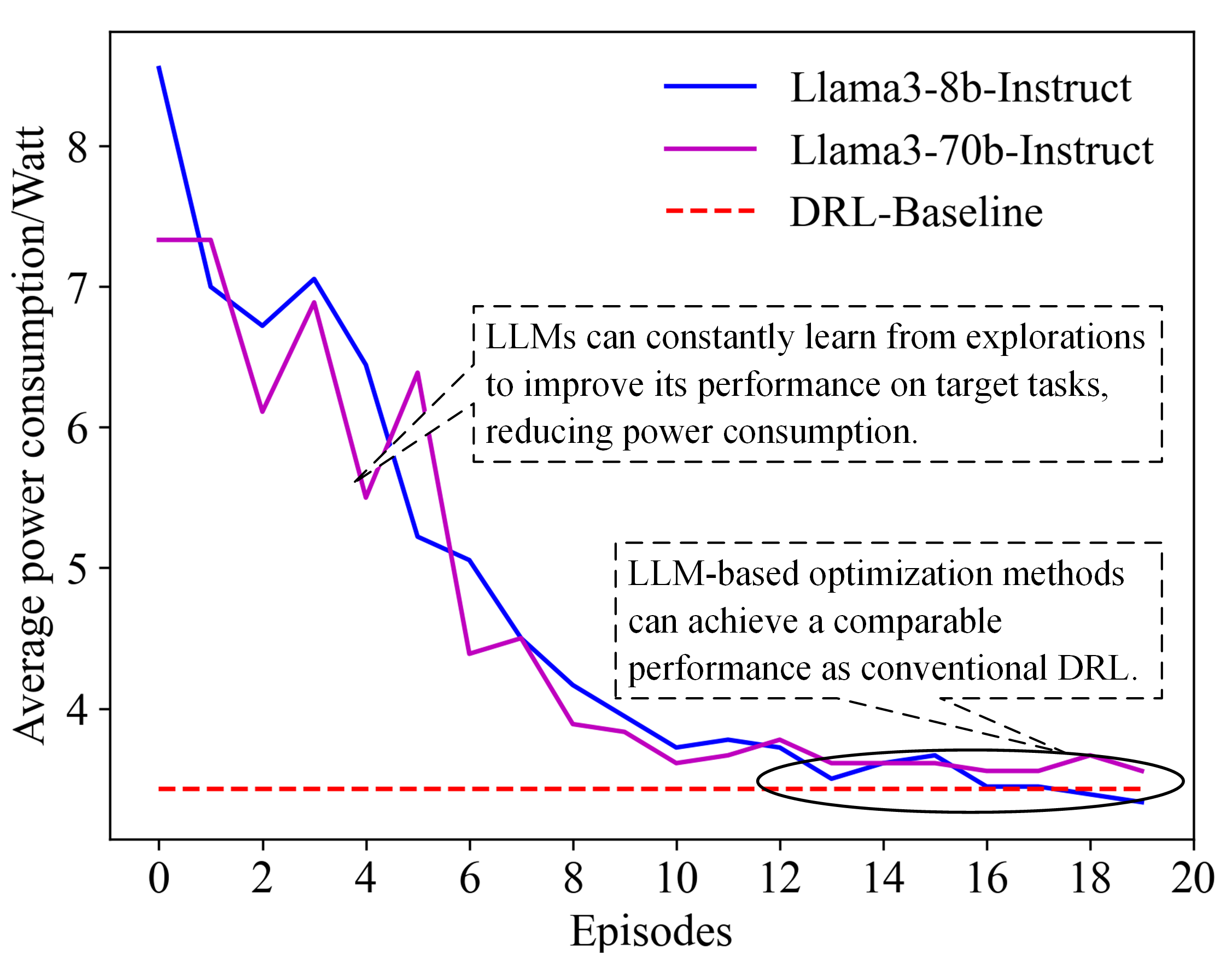}
}
\subfigure[\textbf{Optimization Case Study}: service quality comparison with training.]{ \label{fig_quality}
\includegraphics[width=0.46\linewidth]{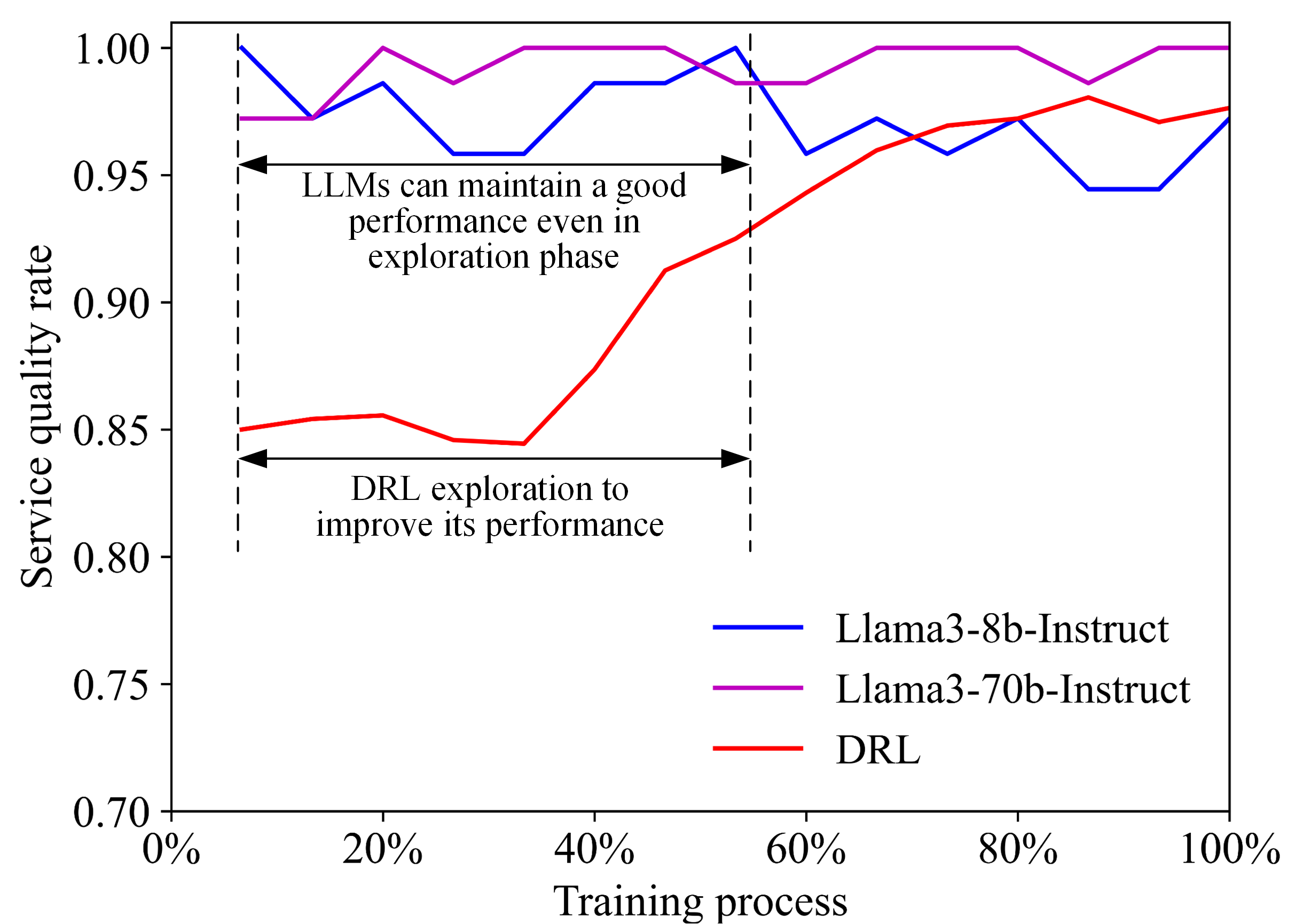}
}
\subfigure[\textbf{Prediction Case Study}: average MAE and MSE comparison. The self-refine method achieves comparable performance as LSTM as a few-shot predictor, while LSTM has been delicately trained on the dataset. 
]{ \label{fig_mae}
\includegraphics[width=0.47\linewidth]{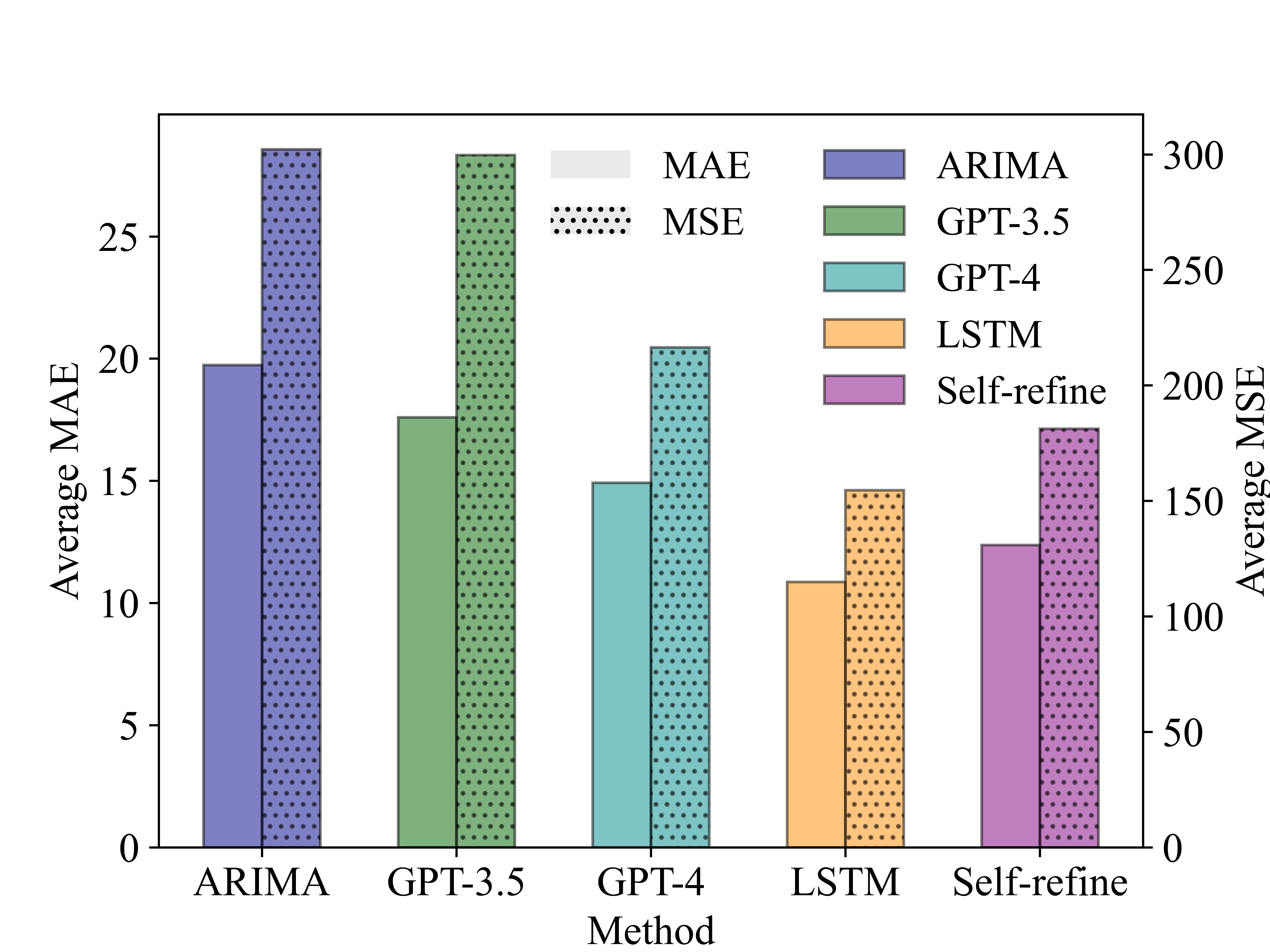}
}
\subfigure[\textbf{Prediction Case Study}: predicted traffic comparison in 24 hours.]{  \label{fig_traffic}
\includegraphics[width=0.48\linewidth]{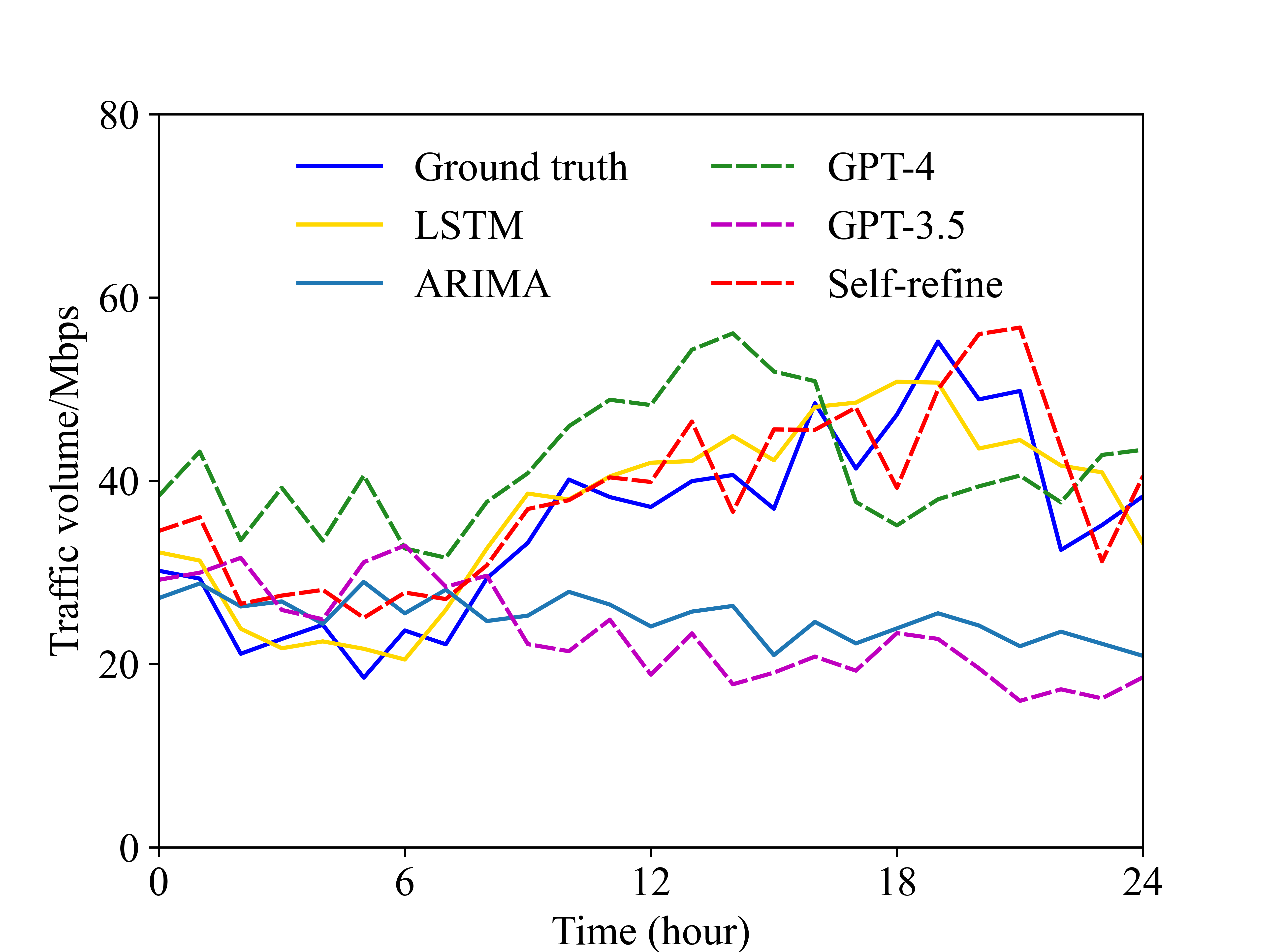}
}
\caption{Performance comparison in network optimization and traffic prediction case studies.}\label{fig-resu}
\end{figure*}

\section{Case Studies}

\subsection{Simulation Settings}
This section presents two case studies on network optimization and prediction problems. \\
1) \textbf{Network Optimization}: This case study involves a base station power control problem a fundamental optimization task in wireless network fields. 
\blue{The considered problem formulations aim to minimize the base station power consumption and meanwhile maintain an average data rate threshold for users}\footnote{\blue{Detailed equations can be found in \cite{chiang2008power} (equation (1.12) in Section 1.3.5)}}.
We consider three adjacent base stations and the associated user numbers dynamically change from 5 to 15 each, the average data rate threshold is 1.5 Mbps/per user, and the channel gain applies 3GPP urban network models.
This case study includes: 
\textbf{\textcircled{1} Iterative prompting}: We consider 2 LLM models: Llama3-7b-instruct as a small-scale model, and Llama3-70b-instruct as a large-scale model. \blue{We deploy the proposed iterative prompting technique as introduced in Fig. \ref{fig-optimization} and Section \ref{sec-opti}, in which the LLM will explore the environment, accumulate experience, and learn iteratively.}
\textbf{\textcircled{2} DRL baseline}: DRL is included as a baseline, which has been widely used to address network optimization problems after dedicated model training. \blue{We apply a classic deep Q-learning algorithm, in which the neural networks have 3 layers, the experience pool size is 10000, the batch size is 64, and the learning rate is 0.005.} 

2) \textbf{Network Prediction}: 
The evaluation dataset is a publicly available network traffic dataset from the city of Milan~\cite{barlacchi2015multi}, recording the time of user interactions and the base stations managing these interactions. 
\blue{The original grid of $100\times 100$ base stations is reorganized into $25\times 25=625$ aggregated base stations. Each aggregated base station covers an area of approximately $1 km\times 1km$, including 8,923 samples. For the evaluation, we randomly select one base station, partitioning the dataset such that 70\% is used for training, 10\% for validation, and the remaining 20\% for testing.} 
The simulation algorithms include:
\textbf{\textcircled{1} Self-refinement}: Our proposed self-refined prompting utilizes GPT-4 as the foundation model for one-day-ahead traffic prediction using historical traffic from the previous day. \blue{It follows the proposed self-refined prompting techniques as defined in Fig.~\ref{fig_self_refine} and Section \ref{sec-pre}.} 
\textbf{\textcircled{2} LLM-related baselines}: We include GPT-4 and GPT-3.5 without self-refinement as two LLM-related baselines, better demonstrating the capabilities of self-refinement. 
\textbf{\textcircled{3} Conventional baselines}: Autoregressive integrated moving average (ARIMA) and LSTM as two conventional baselines.
\blue{Through extensive experiments, the optimal parameters for the ARIMA baseline model are determined including autoregressive ($p=2$), differences ($d=1$), and moving average ($q=2$) components.}
We deploy LSTM as an optimal baseline, since it has been specifically trained on the target network traffic dataset. By contrast, \blue{\textbf{Self-refinement}} and  \textbf{LLM-related baselines} have no previous knowledge on the target task.

\subsection{Simulation Results and Analyses}

\textbf{1) Network Optimization Case Study}: 
Fig. \ref{fig_power} and \ref{fig_quality} present the optimization task results. Specifically, one can observe that the average power consumption is constantly improved with the increasing number of episodes in Fig. \ref{fig_power}. 
It indicates that the proposed scheme in Section \ref{sec-opti} enables LLMs to learn from previous examples and experience, and then improve its performance on target tasks iteratively.
Fig. \ref{fig_power} also shows that LLMs present comparable performance as the DRL algorithm, which is achieved without model training or fine-tuning as in conventional ML algorithms. 
Meanwhile, Fig. \ref{fig_quality} compares the service quality of Llama3-7b, Llama3-70b, and DRL, which means the probability of violating the preset average data rate threshold of 1.5 Mbps/per user. 
It highlights that LLMs can maintain high service quality at the beginning of the training process, while the DRL algorithm has a lower service quality in the initial exploration phase.
\blue{Fig. \ref{fig_quality} reveals that LLMs have a higher learning efficiency than the DRL algorithm. It can be explained by LLM's few-shot learning capabilities, in which LLMs can quickly learn from demonstrations and task descriptions to improve decisions. 
By contrast, the DRL approach relies on exploration and value updating to improve the policy gradually, and the low sampling efficiency prevents the learning efficiency.}

\textbf{2) Network Prediction Case Study}: 
Fig.~\ref{fig_mae} compares the average MAE and MSE metrics of various techniques.
It is observed that the self-refined prompting method significantly outperforms ARIMA, GPT-3.5, and GPT-4, demonstrating the capabilities of our proposed self-refined prediction method.
The self-refined prompting method exhibits notable improvements over GPT-3.5 and GPT-4, with MAE reductions of 29.72\% and 17.09\%, which underscores more powerful generalization capability on unseen datasets through iterative feedback generation and prediction refinement processes.
GPT-3.5 shows a worse performance due to its outdated architecture and designs, which have been observed in many existing studies.
\blue{Although LSTM serves as an optimal baseline with superior performance, it is important to note that LSTM is specifically trained on the target dataset, incurring additional computational costs, which presents significant challenges in scenarios involving resource-constrained devices within wireless networks. 
Given the dynamic evolution of traffic distribution in non-stationary wireless networks, a well-trained LSTM may overfit the historical traffic used during the pre-training, resulting in degraded performance on new, unseen traffic over time.} 
Furthermore, similar results can be observed in Fig.~\ref{fig_traffic}, which illustrates the one-day ground truth traffic and the predicted traffic of various methods.   
The self-refined prediction method more closely matches the ground truth traffic, indicating strong scalability of predictions across various base stations. By contrast, GPT-4 shows an obvious mismatch from 9:00 to 15:00.

In summary, the simulation results in Fig. \ref{fig-resu} demonstrate that LLMs have great potential in handling network management, i.e., optimization and prediction tasks. 
They present satisfied performance as conventional machine learning algorithms such as DRL and LSTM, while avoiding extra model training and fine-tuning complexity. 
LLMs also show higher learning efficiency than existing techniques, and they allow human language-based input, showing the possibility of natural language-based future network management.